\documentclass[journal=ancac3,manuscript=article]{achemso}

\usepackage{chemformula} 
\usepackage[T1]{fontenc} 
\DeclareUnicodeCharacter{2212}{\textendash}
\usepackage{siunitx}

\author{Jonathan T. Weber}
\affiliation{Institute of Physics, Carl-von-Ossietzky University of Oldenburg, 26129 Oldenburg, Germany}
\alsoaffiliation[]{Department of Physics, University of Regensburg, 93053 Regensburg, Germany}

\author{Niklas Müller}
\affiliation{Institute of Physics, Carl-von-Ossietzky University of Oldenburg, 26129 Oldenburg, Germany}
\alsoaffiliation[]{Department of Physics, University of Regensburg, 93053 Regensburg, Germany}

\author{Alexander Schröder}
\affiliation[]{Department of Physics, University of Regensburg, 93053 Regensburg, Germany}

\author{Sascha Schäfer}
\affiliation[]{Department of Physics, University of Regensburg, 93053 Regensburg, Germany}
\alsoaffiliation[]{Regensburg Center for Ultrafast Nanoscopy (RUN) University of Regensburg, 93053 Regensburg, Germany}
\email{sascha.schaefer@ur.de}

\title[]{Visualizing Standing Light Waves in Continuous-Beam Transmission Electron Microscopy}

\begin{document}

\section{Abstract}
The phase-resolved imaging of confined light fields by homodyne detection is a cornerstone of metrology in nano-optics and photonics, but its application in electron microscopy has been limited so far. Here, we report the mapping of optical modes in a waveguide structure by illumination with femtosecond light pulses in a continuous-beam transmission electron microscope. Multi-photon photoemission results in a remanent charging pattern which we image by Lorentz microscopy. The resulting image contrast is linked to the intensity distribution of the standing light wave and quantitatively described within an analytical model. The robustness of the approach is showcased in a wider parameter range and more complex sample geometries including micro- and nanostructures. We discuss further applications of light-interference-based charging for electron microscopy with in-situ optical excitation, laying the foundation for advanced measurement schemes for the phase-resolved imaging of propagating light fields.

\section{Introduction}
Confined light fields enable advanced imaging and sensing methodologies with nanoscale spatial resolution and enhanced sensitivity \cite{novotny_principles_2006}. Homodyne detection, as an established phase-resolved imaging approach, involves the superposition of a light field with a reference wave of equal frequency thereby creating standing wave patterns. Tuning the delay of the reference wave allows for a mapping of the confined light field at different phases. Examples for such approaches within the field of nanoscopies include the various forms of scattering-type scanning near-field optical microscopies (s-SNOM), where the scattered light field from a metal tip interferes with an optical reference wave, enhancing the imaging resolution far beyond the diffraction limit due to the light confinement at the tip apex and the field-sensitivity by homodyne detection \cite{zenhausern_scanning_1995,knoll_near-field_1999,eisele_ultrafast_2014,esmann_vectorial_2019,mooshammer_quantifying_2020}. 
Two-photon photoemission electron microscopy (2PPE-PEEM) serves as another example, in which a first light pulse excites  partially confined propagating wave packages and a second light pulse interferes with the temporally-evolved optical field. At the crests of the interference pattern non-linear photoemission occurs and the resulting photocurrent is spatially mapped by the imaging optics of the photoelectron microscope \cite{kahl_normal-incidence_2014,lemke_mapping_2012}. Both s-SNOM and 2PPE-PEEM have enabled the imaging of transient hybrid light-matter quasiparticles \cite{rang_optical_2008,dai_tunable_2014,spektor_revealing_2017,kahl_direct_2018,jauk_photoemission_2024,frank_short-range_2017,kostina_optical_2019} and light fields with topological character \cite{davis_ultrafast_2020,ghosh_spin_2023}. Also for single pulse-excitation, 2PPE-PEEM was utilized to image the standing wave patterns in waveguide geometries \cite{fitzgerald_subwavelength_2014}. 

%


In transmission electron microscopy (TEM), the intensity distribution of confined light fields are routinely imaged by low-loss electron energy spectroscopy \cite{garcia_de_abajo_optical_2010,polman_electron-beam_2019,hyun_measuring_2008,nelayah_mapping_2007,talebi_wedge_2016,colliex_electron_2016,maciel-escudero_probing_2023}, utilizing the superb spatial resolution of electron microscopy and the recently achieved meV-spectral resolution in high-end microscopes \cite{krivanek_vibrational_2014,li_three-dimensional_2021,qi_measuring_2021}. The emergence of ultrafast transmission electron microscopy (UTEM) \cite{flannigan_4d_2012,feist_ultrafast_2017,houdellier_development_2018,bucker_electron_2016,mattes_femtosecond_2024,gnabasik_imaging_2022,madan_charge_2023,kim_light-induced_2020,zhu_development_2020,olshin_characterization_2020} has additionally enabled the local mapping of the inelastic electron-light scattering efficiency, resulting in a broadly applicable method termed photon-induced near-field electron microscopy (PINEM) \cite{barwick_photon-induced_2009,garcia_de_abajo_multiphoton_2010,park_photon-induced_2010,piazza_simultaneous_2015,feist_quantum_2015,vanacore_ultrafast_2019,wang_coherent_2020,nabben_attosecond_2023,bucher_coherently_2024,muller_inelastic_2024}.
Whereas traditional PINEM is sensitive to certain Fourier components of the intensity distribution of optical near-fields, similar to EELS \cite{liebtrau_spontaneous_2021}, the combination of phase-modulated electron pulses with PINEM has given rise to a phase-resolved mapping of optical near-fields \cite{ryabov_attosecond_2020, gaida_attosecond_2024,bucher_free-electron_2023}. Despite the tremendous progress in these research fields, the much simpler concept of using light interference for the advanced imaging and analysis of nano-optical modes in electron microscopy is so-far only scarcely applied. Examples include the scattering of electrons off standing light waves in Kapitza-Dirac-like experiments \cite{kapitza_reflection_1933,freimund_observation_2001,kozak_inelastic_2018} which can be even utilized as laser-based phase plates for electron beams \cite{schwartz_laser_2019,chirita_mihaila_transverse_2022}, or the interference of cathodoluminescence light \cite{losquin_link_2015,talebi_spectral_2016} in double interaction geometries \cite{taleb_phase-locked_2023}. \\ 
Here, we demonstrate the visualization of standing light waves in a continuous-beam transmission electron microscope with in-situ optical illumination. Specifically, we show that a spatially structured femtosecond (fs) light field leaves a remanent, multi-photon-photoemission-induced charging pattern in a thin aluminum\slash aluminum-oxide layer which can be imaged by in-situ Lorentz microscopy. The versatility of the approach is showcased by visualizing the inhomogeneous distribution of light at edges and in the vicinity of metallic micro- and nanostructures.

\section{Results and Discussion}
As a model system to investigate the emerging Lorentz contrast induced by standing light waves in transmission electron microscopy, we firstly consider specimens with a sharp one-dimensional discontinuity in the refractive index. 
Specifically, we utilize a 50-nm thick silicon nitride (Si$_3$N$_4$) membrane covered with a 4~nm aluminum film. Due to exposure to atmosphere, the aluminum film oxidizes \cite{gorobez_growth_2021,campbell_dynamics_1999}, forming a layer with strongly reduced electrical conductivity. This electron-transparent heterostructure is supported by a 200-µm thick silicon frame with etched quadratic windows of 100-µm length, which effectively scatters the incident light. In all experiments, we applied a sample tilt of 20° unless stated otherwise.

For the sensitive detection of photoemission-induced charge build-up in the specimen, we perform Lorentz-mode electron microscopy with a large defocus of the imaging system (D$_{\mathrm{z}}$~=~-7.5~mm), enabling the imaging of phase shifts of the incident electrons due to the electrostatic fields in the sample region \cite{phatak_recent_2016}.

Illuminating the edge of the silicon frame with s-polarized ultrashort light pulses (central wavelength: $\lambda=800$~nm, pulse duration: 170~fs, estimated focal spot size: 17~µm, repetition rate: 400~kHz), a strong modulation of the electron density on the detector emerges with a periodic pattern of bright and dark lines parallel to the edge of the silicon frame, as shown in Fig.~1a. An analysis for illumination with p-polarized light is given in the Supporting Information (SI).

\begin{figure}
  \includegraphics[scale=1]{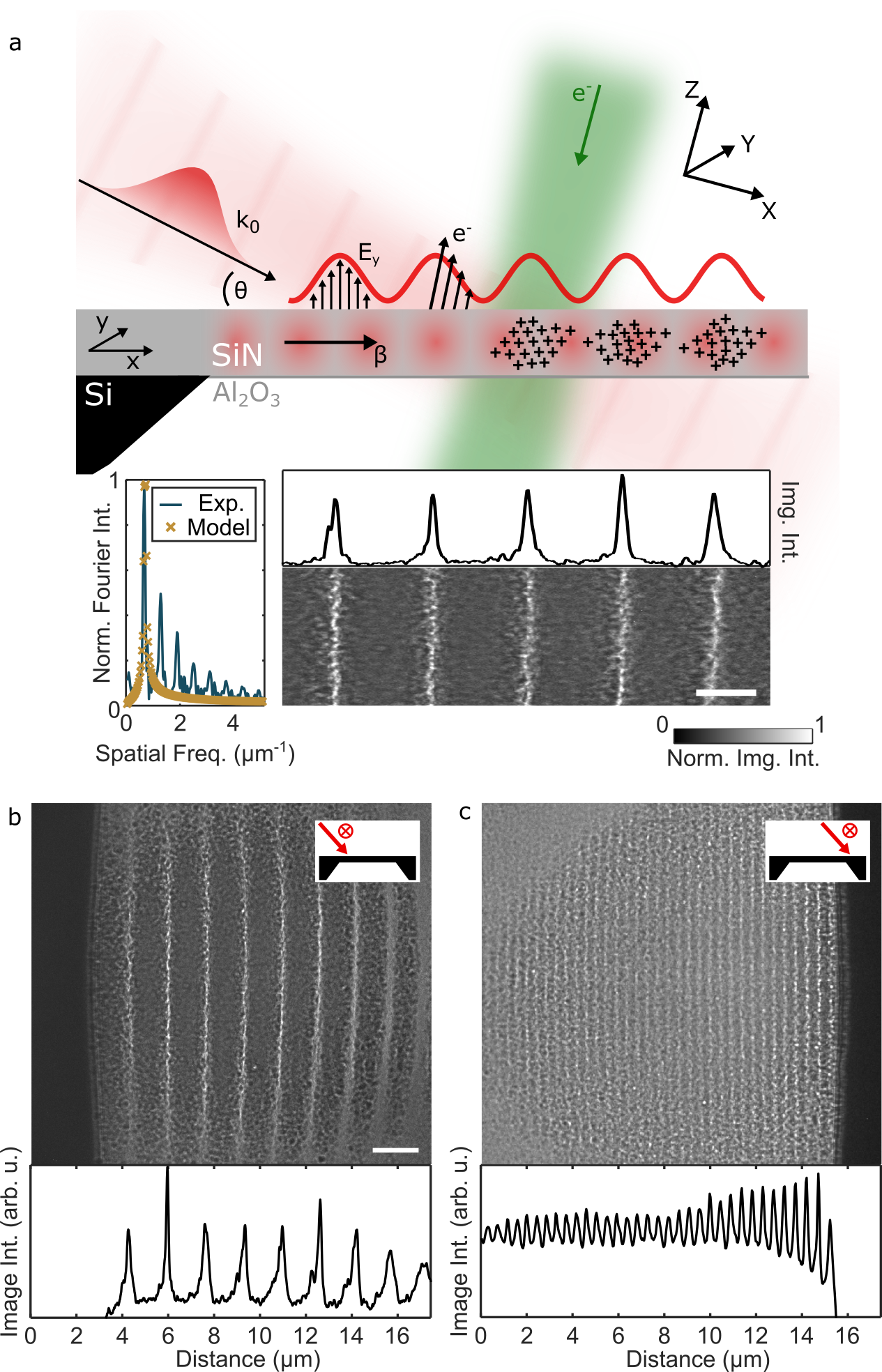}
 \caption{Standing optical wave induced charge patterns in electron microscopy. (a) Scheme of the experimental setup. Incident femtosecond light pulses trigger guided waves within thin film samples by scattering at the edge of a silicon frame. Interference between the guided wave and the incident light forms standing waves which drive non-linear photoemission at the intensity maxima. Under out-of-focus imaging conditions in transmission electron microscopy this light-driven charging mechanisms give rise to strong periodic image contrast modulations. The coordinate system shown in the sketch is rotated to account for the sample tilt. (b and c) Experimental Lorentz micrographs of the resulting charging patterns for illumination of opposite sides of the silicon frame, along with the integrated image intensity, showing the periodic contrast modulations. The sketches indicate the illumination conditions. Scale bar: 2~µm. The inset in (a) shows a cutout of the experimental micrograph (scale bar: 1~µm), along with the Fourier transform of the line profile extracted from the micrograph in (b), and the expected Fourier intensity trace from the standing-wave model.}
  \label{fgr:1}
\end{figure}

Upon illumination of the opposite edge of the window in the silicon frame, we again obtain a stripe-like image contrast, albeit with a higher spatial frequency of the pattern (see Fig.~1c). In both cases, the pattern contrast decreases with larger distances from the edge. Turning off the illumination, the image contrast again becomes homogeneous, ruling out light-induced irreversible sample damages. Furthermore, the modulated image contrast under illumination is only visible in defocused micrographs, demonstrating that the origin of the contrast is related to an imprinted phase on the imaging electron wave. We attribute this phase change to spatially inhomogeneous charging patterns on the sample induced by standing light waves. \\
To calculate the expected pattern of the standing optical wave on the sample, we model the incident light as a plane wave with wavevector $\vec{k}$, which lies in the (x,z)-plane and forms an angle of about 57° with the electron beam (Z-axis). As an optical waveguide, the Si$_3$N$_4$~-~Al$_2$O$_3$ heterostructure supports a TE$_0$ mode with an effective propagation constant of $\beta$~=~9.05~µm$^{-1}$ (see Materials and Methods for details on the multilayer waveguide dispersion characteristics)\cite{chen_foundations_2006}. The edge of the silicon frame acts as a scattering source and couples light into the waveguide structure.  The interference of the incident and the guided light within the heterostructure is governed by the difference of the surface-projected wavenumbers $k_{||}$-$\beta$, with $k_{||} = |\vec{k}|\cos (\theta)$ being the in-plane wavenumber for the incident wave, where $\theta$ is the incidence angle relative to the (x,y)-plane, and $\beta = |\vec{k}| N_{\mathrm{eff}}$ for the guided mode, where $N_{\mathrm{eff}}$ is the effective refractive index for the TE$_0$ mode. The resulting intensity profile of the superposition can be calculated as $I_{\mathrm{res}} = |\Psi_{\mathrm{res}}|^2 \propto 1 + \cos [(\beta-k_{||}) x ]$, leading to a spatial periodicity of the standing wave of $1.9 \times \lambda$ for the adopted experimental parameters. Fig.~1a(inset) shows a comparison between the Fourier transformed line-profile shown in Fig.~1b and the expected trace from our standing-wave model. We extract an experimental peak spatial frequency of 0.6615~µm$^{-1}$ (taking into account the sample-tilt and electron-lensing related image distortions as discussed below), in good agreement with the expected value of 0.659~µm$^{-1}$. If the opposing edge of the silicon frame is illuminated, the expected standing wave period changes to $0.56 \times \lambda$ due to the counter-propagating incident and guided waves, as also observed experimentally (Fig.~1c). In the following, we will focus on the case with lower spatial frequency. \\

\begin{figure}
  \includegraphics[scale=1]{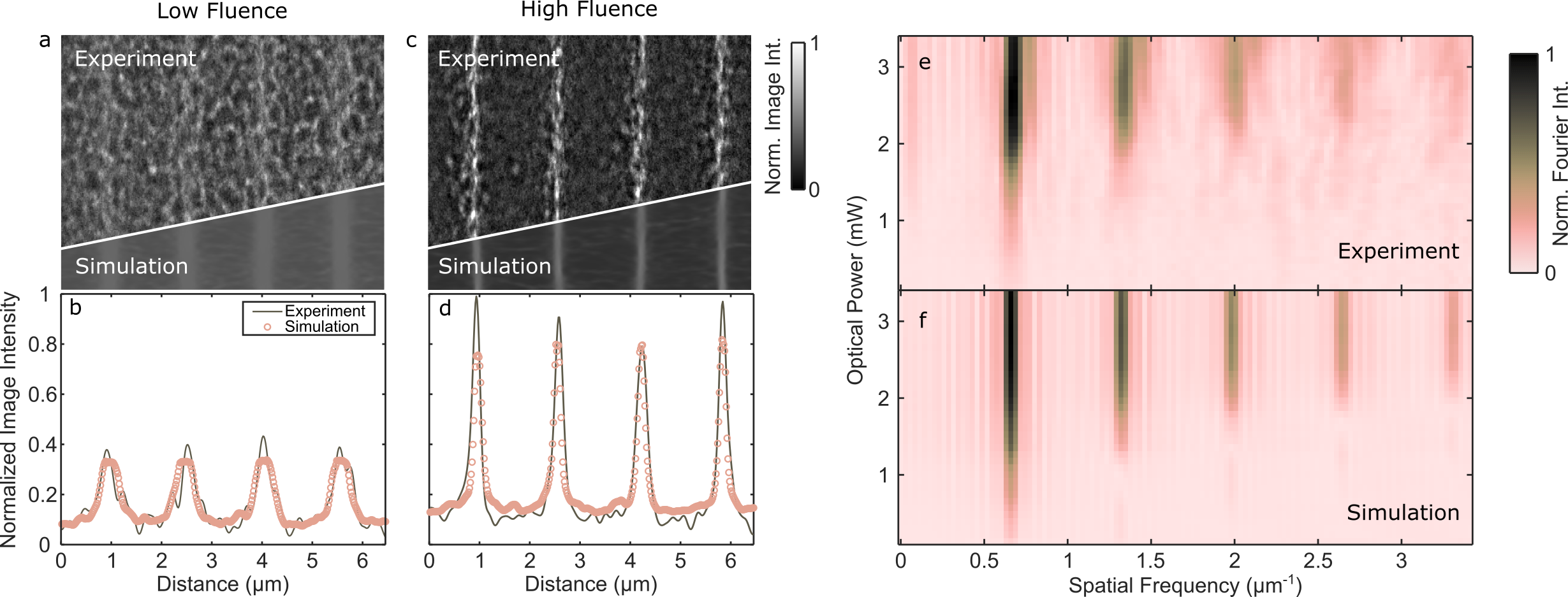}
  \caption{Lorentz image simulations and optical power dependency. (a) Experimental Lorenz micrograph showing the charging pattern induced by optical illumination with an incident optical power of 0.98~mW. The lower part shows the computed Lorentz contrast, expected for multi-photon-photoemission-driven charging of a granular specimen. (b) Comparison of the image intensity integrated along the vertical axis for the experimental and the simulated Lorentz micrograph. (c,d) Same as (a and b) but for an incident optical power of 2.3~mW. (e,f) Fourier transform traces of experimental/simulated image intensities integrated along the vertical axis for varying incident optical powers. The simulated image intensities are determined by a non-linear least squares fitting to the experimental micrographs, taking into account light-induced decharging mechanisms by a resizing factor \cite{crameri_misuse_2020}.}
  \label{fgr:2}
\end{figure}

To study the influence of the optical power on the resulting charging patterns, we control the incident fluence by a variable neutral density filter, while recording electron micrographs in Lorentz mode. We observe an increased image contrast for higher optical powers and the periodic contrast variation transitions from a more sinusoidal pattern (Fig.~2(a,b), 0.98~mW incident power) to a structure characterized by distinct bright lines (Fig.~2(c,d), 2.34~mW incident power) in agreement with a sharpening of the corresponding line profiles in Fig.~2(b,d) (dark lines). Correspondingly, Fourier transforms perpendicular to the stripe pattern for varying incident optical power, as shown in Fig.~2e, reveal an increased contribution from higher spatial harmonics at large optical powers. Generally, such a behaviour can result from the inherent nonlinearity of the photoemission process but also from the nonlinear contrast mechanism operative at large defoci (for experimental defocus scans, see SI). We reproduce the experimental micrographs by Lorentz image simulations assuming the charge distribution in the sample to follow the cube of the local intensity, i.e. $I_{\mathrm{res}}^3$, as expected for a three-photon photoemission processes. From this two-dimensional charge-distribution $\rho (x,y)$, the resulting inhomogeneous phase-shift of the imaging electron wave $\phi (x,y)$ is calculated by solving 
\begin{align}
  \nabla ^2 \phi (x,y) \propto - \rho (x,y).
\end{align}

For details of the Lorentz image simulations, see the SI. We fit the simulated Lorentz contrast to the experimental micrographs by minimizing the squared differences between the image intensities, having the induced photovoltage and the relative phase of the standing wave as free parameters. The computed results are shown in the lower part of Fig.~2(a,c) and the corresponding integrated line profiles are presented in Fig.~2(b,d) (pink circles). 

Due to an overall charging of the imaged sample region and the resulting electron lensing effect, we observe an optical-power-dependent effective magnification \cite{weber_electron_2024}, which changes by a factor of 1.1 between 0.13~mW and 3.36~mW, as discussed in the SI. The spatial scalings in Fig.~2 were adapted utilizing this effective magnification.

\begin{figure}
  \includegraphics[scale=1]{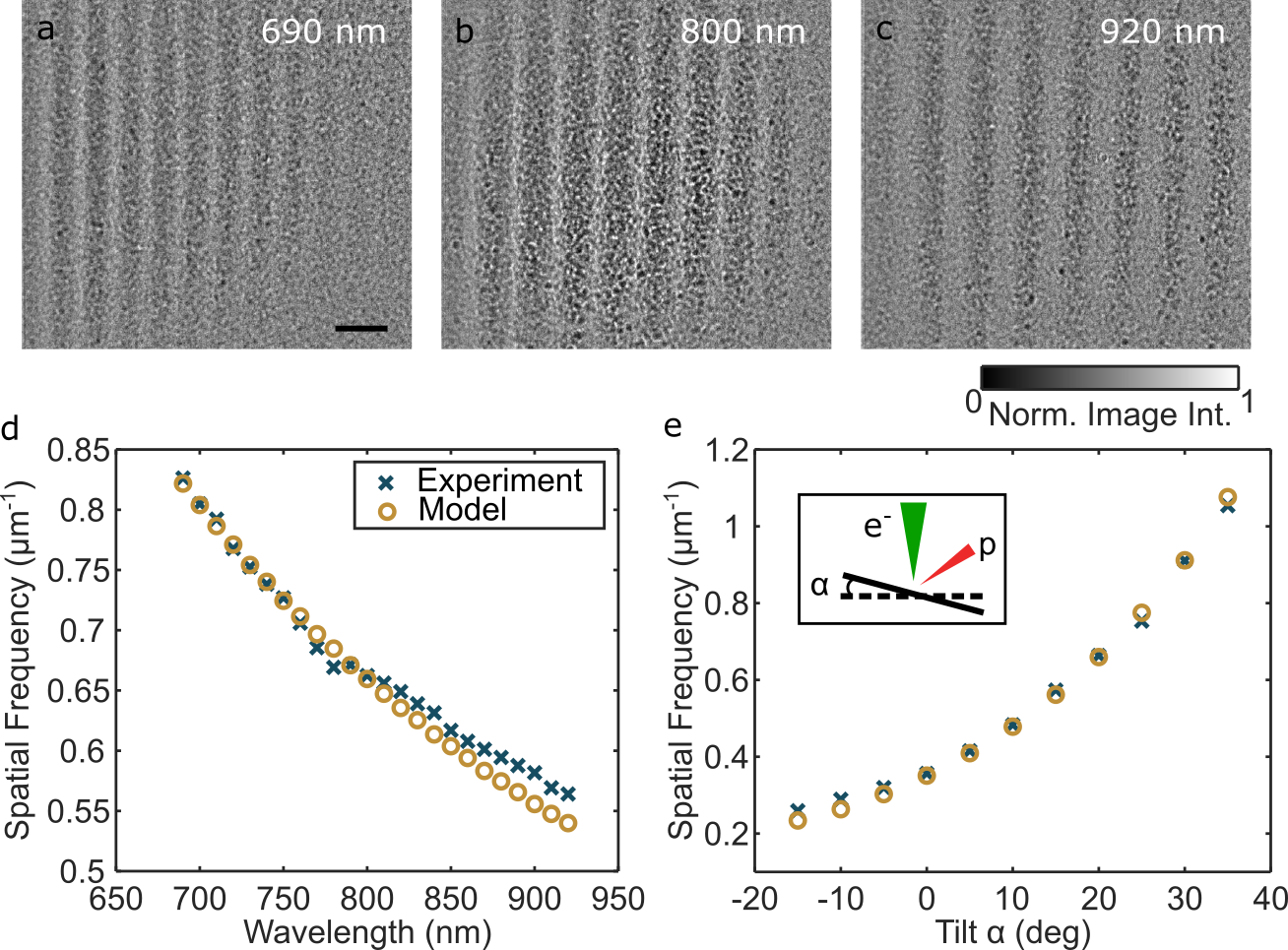}
 \caption{Effect of incident optical wavelength and sample tilt on the observed charging pattern. (a--c) Experimental micrographs showing the resulting image intensity modulations for incident optical wavelengths of 690~nm, 800~nm and 920~nm, respectively. Scale bar: 2~µm.(d) Extracted spatial frequencies for varying incident optical wavelengths, along with the expected spatial frequencies according to the standing wave model discussed in the text. (e) Same as in (d) but for a varying tilt angle $\alpha$ of the sample in the electron microscope.}
  \label{fgr:3}
\end{figure}

As a next step, we investigated the influence of the photon energy on the charging process, recording electron micrographs of the charge pattern induced by the standing optical wave while varying the incident optical wavelength from 690~nm to 910~nm (at a fixed incident optical power of 1 mW). In Fig.~3(a-c) exemplary Lorentz micrographs at three indicated excitation wavelength are presented, showing a shift towards lower spatial periodicities for longer optical wavelengths, as expected from the standing-wave model. We extracted the peak position of the Fourier-transformed Lorentz micrographs along the x-axis, and plotted them in Fig.~3d together with the expected spatial frequency from our model. 

In Fig.~3e, the extracted peak spatial periodicities (blue crosses) are presented for a sample tilt from $\alpha$~=~$-15$° to $35$°, where 0° indicates that the sample normal is parallel to the incident high-energy electrons (Z-axis). The according micrographs were recorded with an incident optical wavelength of 800~nm at 1~mW power and subsequently resized by the appropriate factor, also taking into account the sample tilt-related image distortions. The experimentally observed periodicities for the wavelength- as well as for the tilt-scan are well reproduced by the standing wave model (yellow circles). \\

\begin{figure}
  \includegraphics[scale=1]{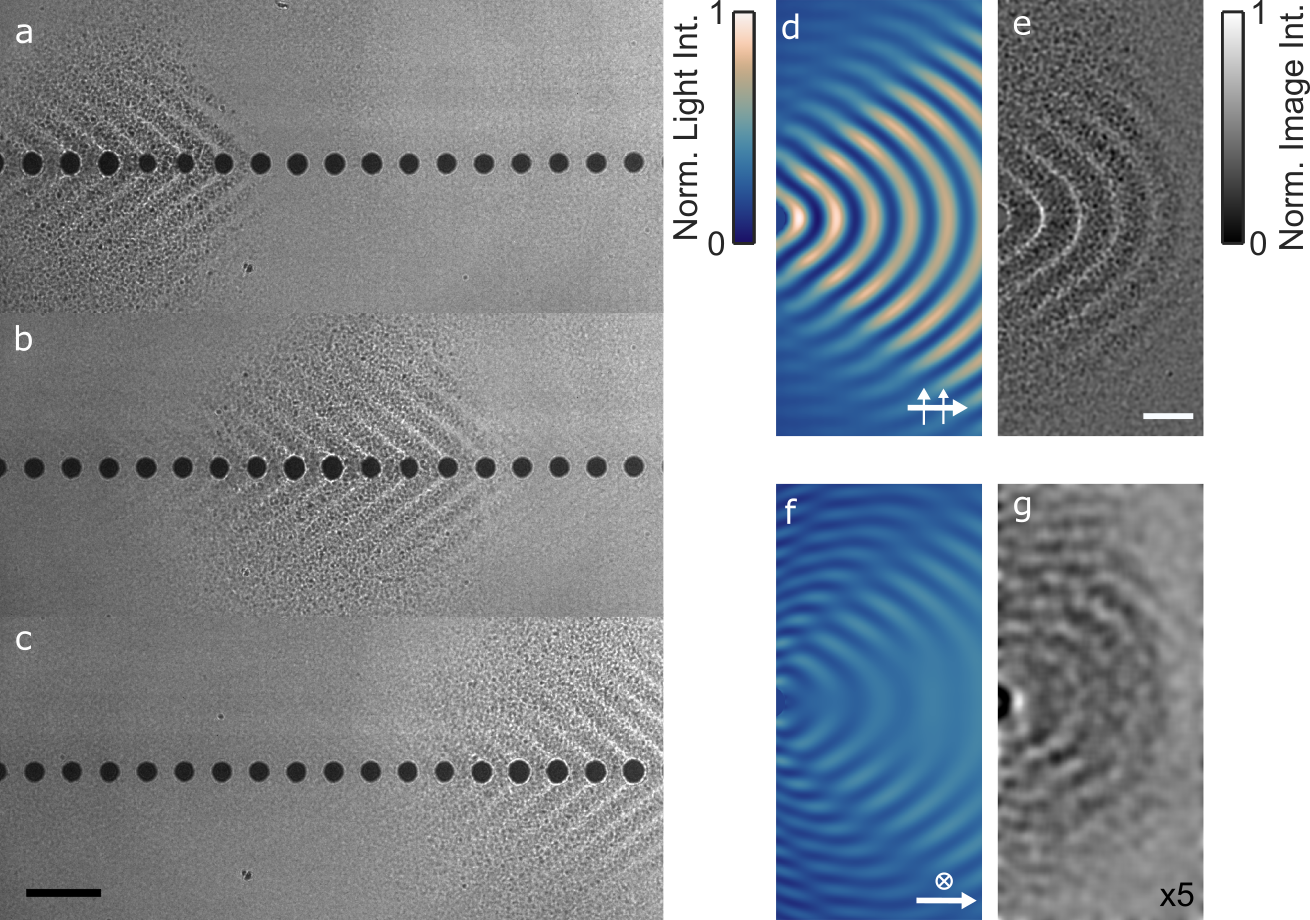}
 \caption{Standing optical waves on nanostructured surfaces and influence of light polarization. (a--c) Experimental Lorentz micrographs of charge patterns resulting from standing optical waves formed by illumination of a linear array of circular gold nanoislands. The optical focus position is moved is moved in the horizontal direction by means of a focusing lens mounted to a piezo stage. Scale bar: 4~µm. (d) Numerically computed light intensity, resulting from illuminating a circular gold nanoisland with s-polarized light. Arrows in the lower right corner indicate the in-plane direction and polarization of the incident light. Scale bar: 2~µm. (e) Lorentz micrograph of the standing wave pattern at the end of the disc array corresponding to the simulated light field in (d). (f and g) Same as (d and e) but for incident p-polarized light. The micrograph in (g) is smoothed with a Gaussian kernel of size 12 and five-fold increase in image contrast as compared to (e) for enhanced visibility of the low-contrast pattern.}
  \label{fgr:4}
\end{figure}

To showcase the versatility of our approach for mapping standing wave light fields in more complex scattering geometries, we prepared a one-dimensional array of circular gold islands (1~µm-diameter, height of 20~nm, disc periodicity of 2~µm) by electron-beam lithography on the silicon-nitride-side of the heterostructure. The structure is again illuminated by a focused light beam and we record the light-induced charging patterns in the vicinity of the array depending on the lateral position of the light focal spot. Figure~4(a-c) displays three exemplary micrographs with relative lens shifts of 0~µm, 11~µm, and 21~µm, respectively. 
An assembled movie with further relative shift values is shown in the supporting information. In this geometry, the emerging standing wave pattern consists of periodic stripes at an angle of around $\pm 34.5$°, in good agreement with the expected result based on diffraction from a one-dimensional periodic lattice and interference with the incident wave (see Material and Methods for details).

So far, all presented experiments where conducted with s-polarized incident light for which numerical simulation predicted higher interference contrast for our sample structures. Indeed, imaging the charging pattern at the end of the linear array for s- and p-polarized illumination, respectively, we find a substantially larger image contrast modulation for the s-polarized illumination(Fig.~4e). The corresponding standing wave pattern for p-polarized light is only visible after heavy image smoothing (Fig.~4g, for non-smoothed data see supporting information). An assembled video with intermediate polarization steps is presented in the supporting information. The qualitative features of the standing wave patterns at the end of the linear disc array can be reproduced using the surface electric field numerically computed in a finite-difference time-domain model (see SI), as shown in Fig.~4(d,f). We trace the higher modulation contrast along the horizontal direction in the s-polarized case, which is also found in the simulations, back to the dipolar polarization patterns at the gold discs.

Finally, we want to shortly discuss the putative microscopic charging mechanism which leads to the observed Lorentz contrast \cite{weber_electron_2024}. As visible in Fig.~4(a-c), the illuminated area of the sample not only shows a standing wave contrast pattern but also a discernibly enhanced grainy background structure with occasional charging hot spots. Such image features would be compatible with a multi-photon photoemission from individual metallic aluminum granules, surrounded by a passivating aluminum oxide layer. Furthermore, since the apparent magnification depends on the illumination intensity (SI Fig.~S1f), we conjecture that a significant part of photoemitted low-energy electrons leave the imaged sample area.

We note however that for the current sample system several additional charging processes could be contributing, such as multi-photon photoemission from aluminum oxide \cite{stoian_surface_2002} and silicon nitride and increased charging by light-assisted secondary electron emission, as observed in ultrafast scanning electron microscopy \cite{zani_charge_2018}.

It is notable that we observed low-contrast standing-light-wave-induced charging also on silicon nitride membranes without coating of metallic or dielectric thin films in areas that were exposed to high electron doses. We attribute this to light-induced charging processes in the electron-beam-deposited thin film of amorphous carbon \cite{hettler_carbon_2017-1}. In particular, after oxygen-plasma cleaning of the sample surface, no more light-induced image intensity modulations could be observed in these areas on the sample (see supporting information for Lorentz micrographs). \\

Finally, it is worth mentioning that the approach taken here, in which standing wave patterns are formed by the superposition of incident and scattered wave fields, can be extended in a straightforward manner to a two-pulse excitation scheme, very much akin to two-photon PEEM, which would allow a phase-resolved tracking of propagating light fields in continuous beam transmission electron microscopy. 

\section{Conclusion}
In conclusion we have presented the electron imaging of charge distributions induced by standing optical waves formed by interference of an incident fs-pulse and a guided light mode.
We could successfully describe the charging patterns formed around a one-dimensional discontinuity of the refractive index in the sample plane by an analytical model considering the projection of the incident optical plane-wave onto the sample surface and a guided mode inside the multilayer sample. 
We examined the dependence on several experimental parameters, specifically the excitation fluence, wavelength and sample tilt, which are relevant in the context of transmission electron microscopy with in-situ optical illumination.
For the description of the polarization-dependence of light-induced charging around more complex scattering geometries we performed finite difference time domain simulations. \\
The data presented here is intended to clarify the causes and consequences of optical interference in in-situ electron microscopy and understanding the inhomogeneous distribution of light in samples with complex geometries. We anticipate, that the concepts presented here can be expanded to develop a phase-resolved electron imaging of standing wave patterns based on double-pulse optical excitation. This approach could enable the imaging of propagating confined optical waves in TEM with fs-temporal resolution and swift electrons, thus permitting the investigation of dielectric materials complementary to two-photon PEEM.\\

\section{Acknowledgement}
The authors would like to acknowledge financial support by the Volkswagen Foundation as part of the Lichtenberg Professorship “Ultrafast nanoscale dynamics probed by time-resolved electron imaging”. Additionally, the authors thank the German Science Foundation (DFG) for the funding of the ultrafast transmission electron microscope (INST 184/211 1 FUGG), the electron-beam lithography instrument (INST 184/107-1 FUGG) and the funding within the priority program 1840 ”Quantum Dynamics in Tailored Intense Fields”.
Furthermore, the authors acknowledge support by the Free State of Bavaria through the Lighthouse project "Free-electron states as ultrafast probes for qubit dynamics in solid-state platforms" within the Munich Quantum Valley initiative. 

\section{Material and Methods}

\subsection{Transmission electron microscopy with in-situ femtosecond optical excitation}
All electron micrographs were recorded using the Regensburg ultrafast transmission electron microscope - a modified JEOL JEM-F200 multi-purpose transmission electron microscope equipped with a laser-driven cold field-emitter electron gun and, in this work, operated in a continuous-electron-beam mode at an electron energy of of 200~keV. Lorentz micrographs were recorded in the low magnification mode of the TEM (4k nominal magnification) with a defocus of $-7.5$~mm, unless otherwise stated. The condenser system was adjusted for a homogeneous electron illumination of a large area to minimize electron-beam induced lensing effects, with a spot size of 4 and a 200-µm diameter condenser lens aperture.
For image acquisition we used a complementary metal oxide semiconductor detector (TVIPS TemCam-XF416R, 4096 pixels × 4096~pixels, 15.5-µm pixel size, 4x4 binning, 5-s exposure time). \\
Optical illumination of the sample is enabled by a home-built light-incoupling unit at the height of the pole piece, consisting of a vacuum viewport and a focusing lens mounted to a 3D piezo stage. The light is incident at an angle of about 57° relative to the electron beam. The optical setup is based on an amplified Yb-doped potassium gadolinium tungstate (KGW) femtosecond laser system (Carbide, Light Conversion) and a collinear optical parametric amplifier (OPA, Orpheus HP, Light Conversion). For ensuring a stable optical spot position on the sample, a beam stabilization system (TEM Messtechnik, Aligna) is used with the spatial and angular optical detectors attached to the TEM. 

\subsection{Specimen preparation}

For imaging the light standing-wave patterns at edge-like structures, commercial silicon nitride membranes were used (200-µm silicon frame thickness, 50-nm membrane thickness, supplier: PELCO). The backside of the membrane was coated with a 4~nm thick aluminum film using electron-beam vapor deposition (base pressure of about 10$^{-7}$~mbar). Subsequently the sample was exposed to atmospheric oxygen so that the ultrathin aluminum film becomes oxidized \cite{gorobez_growth_2021,campbell_dynamics_1999}. For the investigation of light-induced charging patterns around more complex scattering geometries, gold nanoislands were prepared on the Si$_3$N$_4$-side of the sample patterning a mask into a poly(methyl methacrylate) resist by electron-beam lithography and subsequent deposition of 2~nm chromium and 20~nm gold by electron-beam vapor deposition. The thin chromium layer promotes adhesion of the gold film to the surface of the membrane.

\subsection{Calculation of the effective propagation constant in a multilayer waveguide}
For describing guided light modes in the Al/Al$_2$O$_3$/Si$_3$N$_4$ sample stack, we model the sample as a four-layer dielectric step-index waveguide, with the light confined in the two inner layers and the outer layers being vacuum. Since the lateral dimensions of the membrane are much larger than the film thickness, the waveguide is approximated as an infinitely extending structure. For the inner layers, we consider a 50-nm thick silicon nitride film (refractive index n = 1.9962)\cite{philipp_optical_1973} and a 4-nm Al$_2$O$_3$ coating (n = 1.76)\cite{malitson_refraction_1962} and neglect contributions from sparse aluminium grains. 

The optical mode dispersion of the waveguide structure can be calculated by solving the wave equations in the corresponding four regions (with refractive indices $n_i$ and thicknesses $h_i$, with $i = 0,1,2,3$, where $n_0 = n_3 = 1$ and $h_0 = h_3 = \infty$), taking into account the continuity conditions at the boundaries and the constraints of having spatially oscillating solutions within the guiding layers and exponentially decaying solutions in the outer layers \cite{shen_guided_2010,fitzgerald_subwavelength_2014,chen_foundations_2006}.
After some algebraic transformations, the following equation is obtained for the fundamental transversal electric (TE$_0$) mode:
\begin{align}
    k_1 h_1 = \mathrm{tan}^{-1} \left( \frac{q_0}{k_1} \right) + \mathrm{tan}^{-1} \left( \frac{q_2}{k_1} \right).
\end{align}
With
\begin{align}
    q_2 = k_2 \tan \left[ \tan^{-1} \left( \frac{q_3}{k_2} \right) - k_2 h_2 \right],
\end{align}
which can be solved for $\beta$, i.e. the longitudinal propagation constant for the guided mode, by established numeric algorithms.
Here $k_1 = \sqrt{n_1^2 k_0^2 - \beta^2}$, $k_2 = \sqrt{n_2^2 k_0^2 - \beta^2}$ and $q_0 = q_3 = \sqrt{\beta^2 - n_0^2k_0^2}$. $k_0$ is the incident wavevector with $k_0 = \frac{2 \pi}{\lambda}$, where $\lambda$ is the vacuum wavelength. We only consider the TE$_0$ mode, as it can be excited at the silicon edge by the incident s-polarized light and higher-order modes are not supported in a thin-film stack of the above mentioned thickness.

\subsection{Calculation of standing wave pattern around a one-dimensional lattice of scattering centers}
To accurately describe the standing wave pattern formed around an array of circular gold islands with periodicity $d$, we model the islands as a one-dimensional lattice of point-like scattering centers. By satisfying the Laue condition $\vec{k}_{||} - \vec{k}_g = n \vec{G} $, where $\vec{k}_{||}$ is the projection of the incident wave onto the sample plane and $n \vec{G}$ is an integer multiple of the reciprocal lattice unit vector with $|\vec{G}| = \frac{2 \pi}{d}$, we identify the scattered wavevectors of the guided TE$_0$ mode, $\vec{k}_g$ for which constructive interference is observed. The time-integrated superposition intensity of these scattered waves with the incident field can then be written as:
\begin{align}
    I_{\mathrm{sup}} &= | \mathrm{e}^{i k_{||} x} + \mathrm{e}^{i (k_{g,x} x + k_{g,y} y)}|^2 \\
    &= 2 + 2\cos \left( k_{g,y} y + (k_{g,x} - k_{||})x \right)
\end{align}
Thus, the emerging standing wave pattern shows an angle of the standing-wave fronts, given as: 
\begin{align}
    \delta &= \pi/2 - \mathrm{tan}^{-1} \left( \frac{k_{g,y}}{k_{g,x} - k_{||}} \right) \\
    &= \pi/2- \mathrm{tan}^{-1} \left( \frac{\sqrt{k_g^2 - (k_0 + nG)^2}}{nG} \right) 
\end{align}
For $n=1$, we obtain $\delta=37.2$° in good agreement with the experimentally determined value of 34.5$^{\circ}$.

\newpage

\bibliography{Bib_Interference_Paper_final}

\newpage

\setcounter{equation}{0}
\setcounter{figure}{0}
\setcounter{table}{0}
\setcounter{page}{1}
\makeatletter
\renewcommand{\theequation}{S\arabic{equation}}
\renewcommand{\thefigure}{S\arabic{figure}}

\noindent\textbf{\large{Supporting Information for: \\
Visualizing Standing Light Waves in Continuous-Beam Transmission Electron Microscopy}}

\vspace{2cm}
\noindent \textbf{This supporting information includes:} \\ 
Additional information on the finite-difference time-domain simulations, the Lorentz image simulations, the modeling of electric-field induced electron phase shifts and the analysis of the imaging system defocus dependency. \\
Figures S1, S2 and S3. \\
Captions for Movies 1-3. \\
\textbf{Other supporting materials for the manuscript:} \\
Movies 1-3.
\newpage

\section{Description of the videos}
1. Assembled video of the data presented in Fig.4~(a-c), with additional relative lens shift values. \\
2. Assembled video of the data presented in Fig.4~(d,f) with additional, intermediate polarization values for the incident light.\\
3. Assembled video showing the lens shift scan between a linear array of gold nanoislands and the edge of a silicon frame.

\section{Finite-difference time-domain simulations of standing light waves}
We performed numerical finite-difference time-domain simulations using the Ansys Lumerical software suite to determine the surface electric field distribution for the gold disc array. 
For the simulation, we employed the finite-difference time-domain solver in a three-dimensional box with dimensions of: 40~µm $\times$ 35~µm $\times$ 6~µm. The boundaries were treated as perfectly matched layers. We placed the silicon nitride and aluminum-oxide thin film (respective thicknesses of 50~nm and 4~nm) and the edge of a silicon pyramid in the simulation region (respective refractive indices were taken from Refs. $^{1,2}$). 
Within the thin-films the meshing was set manually to assure at least 10 cells along the thickness of the silicon nitride layer and not less than fours cells for the aluminum oxide layer. For illuminating the structure we chose a Gaussian light source with a full-width-at-half-maximum of 17~µm, incident at an 38.7° angle relative to the sample normal with the polarization dependent on the experimental setup we aimed to reproduce. The resulting electric fields were retrieved with a frequency-domain monitor.

\section{Modelling the electric-field-induced electron phase shift and Lorentz image simulations}
We model the influence of a charge distribution, localized in the (x,y)-plane on a plane electron wave propagating along the z-axis.
The two-dimensional inhomogeneous phase shift of the imaging electron plane wave due to electrostatic fields $V (x,y,z)$ is calculated as
\begin{align}
    \phi (x,y) = C_E \int_{-\infty} ^{\infty} V (x,y,z) \mathrm{d}z,
\end{align}
where $C_E$ is a constant factor, depending on the acceleration voltage of the electrons (for 200kV: $C_E$~=~7.29x10$^6$~rad~V$^{-1}$m$^{-1}$).$^{3}$ 
Following Poisson's equation $\nabla ^2 V (x,y,z) = - \frac{\rho (x,y,z)}{\epsilon_0}$, we can write 
\begin{align}
    \nabla ^2 \phi (x,y) = -\frac{C_E}{\epsilon_0} \rho_{\mathrm{proj}},
    \label{eq:poisson_phase}
\end{align}
where $\rho_{\mathrm{proj}}$ is the projection of the charge distribution onto the (x,y)-plane, which is equal to $\rho (x,y,0)$ in our case, since the optically induced charge distribution can be treated as being confined to the sample plane.$^{4}$ 
We solve Eq.~\ref{eq:poisson_phase} by performing a Fourier-transform, isolating $\phi (k_x,k_y)$ and subsequent inverse Fourier transformation, which yields
\begin{align}
    \phi (x,y) = \mathrm{\large \textit{FT}\normalsize^{-1}}  \left( \frac{C_E}{\epsilon_0 |\vec{k}|^2} \mathrm{\large \textit{FT} \normalsize} \left( \rho (x,y) \right) \right),
    \label{eq:phase_shift}
\end{align}
where $\vec{k}$ is the wavenumber with the respective in-plane components $k_x$ and $k_y$. \\
To compute the expected image intensities, we discretize the phase shift map calculated by Eq.~\ref{eq:phase_shift} on a 1024x1024 grid with an effective pixel spacing of 17~nm. To account for the sample's grainy structure and the predominance of photoemission-induced charging on metallic aluminum granules, 85~\% of the entries in the charge-distribution map calculated from the standing wave model are set to zero, with the indices of these entries chosen randomly, before calculating the phase shift $\phi (\vec{r})$.
The imaging electron plane wave is then calculated as $\Psi (\vec{r}) = A(\vec{r}) \mathrm{e}^{i \phi (\vec{r})}$, where the electrons trajectory is chosen to be along the Z-axis and $\vec{r}$ denotes positions within the (x,y)-plane. The amplitude of the imaging electron wave, which becomes modulated by high-angle scattering of the electrons within the sample in combination with the small acceptance angle of the imaging system is assumed to be homogeneous all over the membrane and can therefore be neglected. 
For further details on the calculation of the Lorentz image intensity from the phase shift maps, please see the supporting information of Weber and Schäfer (2024).$^{5}$ 

\section{Analysis of the imaging-system defocus dependency}
To reproduce the micrographs of standing-wave patterns at different defocus settings of the imaging system, we employ the same Lorentz contrast simulation algorithm and fit procedure as described in the main text for the optical-power-dependent measurements.
To accommodate the considerable image transformations that go along with changing the imaging system defocus, we pre-process the experimental micrographs, including shifts, rotations and resizing. For this we use the high-contrast edge of the Si frame and the periodicity of the standing-wave pattern as a reference. \\
Notably, an increasing discrepancy between experimentally determined and simulated image intensities is observable for smaller imaging system defoci, indicating that the charge localization is not fully determined by the cube of the local light intensity. Although it is reasonable to assume that the photoinduced charge distribution can be described by the light field intensity and the nonlinearity of the photoemission process, this does not necessarily imply that the same behavior is observed in a time-averaged measurement scheme. Employing electron holography would allow to accurately capture the charge distribution in a quantitative manner.$^{6}$ \\ 
Furthermore, it is notable that the plateauing of the photovoltage in Fig.~S1e coincides with the observation of irreversible sample changes induced by the standing wave.

\begin{figure}
  \includegraphics[scale=1]{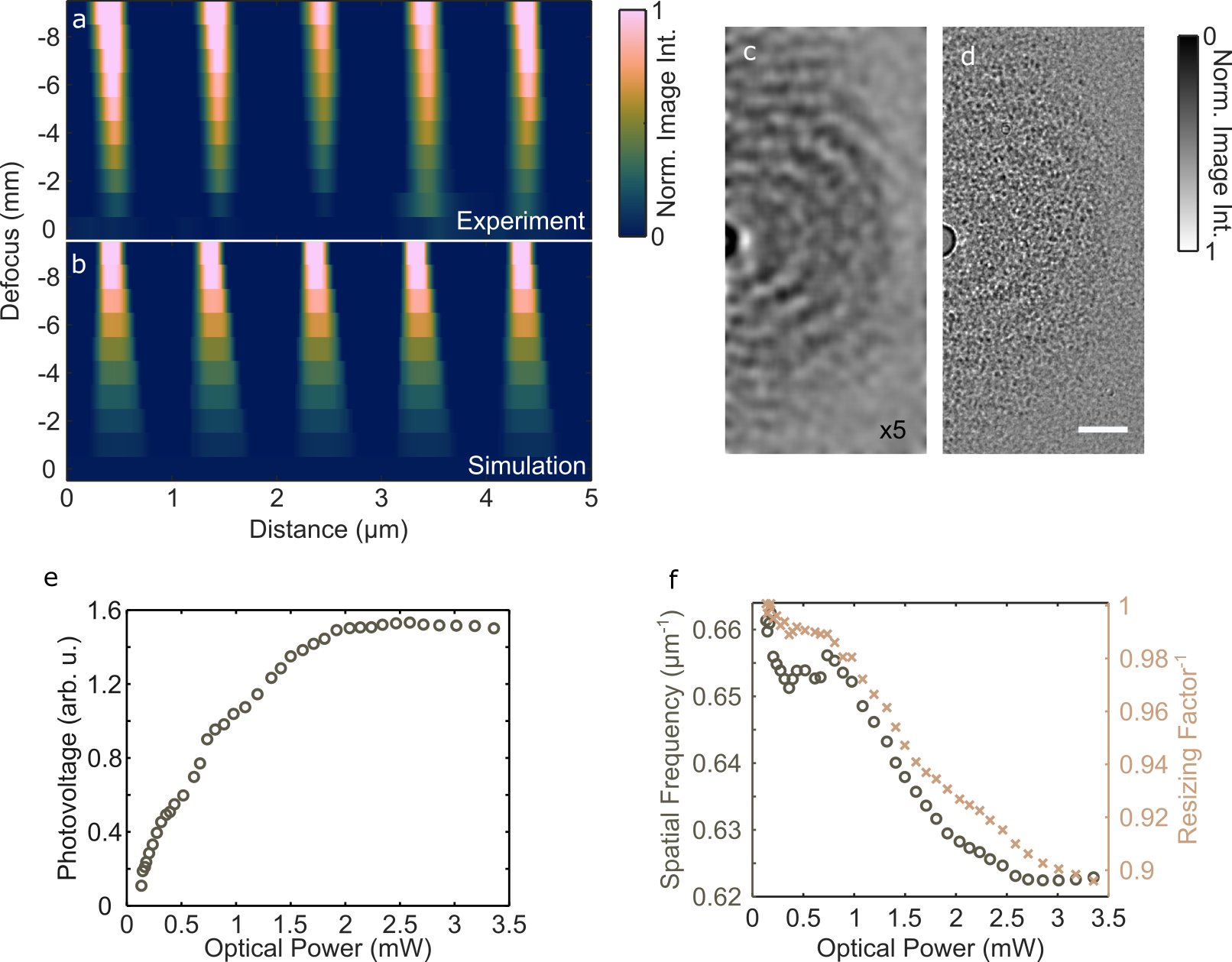}
 \caption{(a) Experimentally retrieved and (b) simulated line profiles showing the standing-wave-induced charging pattern for different defocus settings of the TEMs imaging system. 
 (c) Smoothed and (d) unsmoothed Lorentz micrograph of the standing-wave-induced charge pattern around the end of a linear array of gold nanoislands, under illumination with p-polarized light. To enhance contrast, diminished by the smoothing of the granular background, the image intensity in (c) is multiplied by a factor of 5. Scale bar is 2~µm.
 (e) Results for the photovoltage values from the fit procedure for varying optical fluences as described in the main text. (f) Results for the extraced spatial periodicities of the standing-wave-induced charging pattern for different incident optical fluences along with the inverse of the applied resizing factors from the fit procedure. }
  \label{fgr:S1}
\end{figure}

\begin{figure}
  \includegraphics[scale=1]{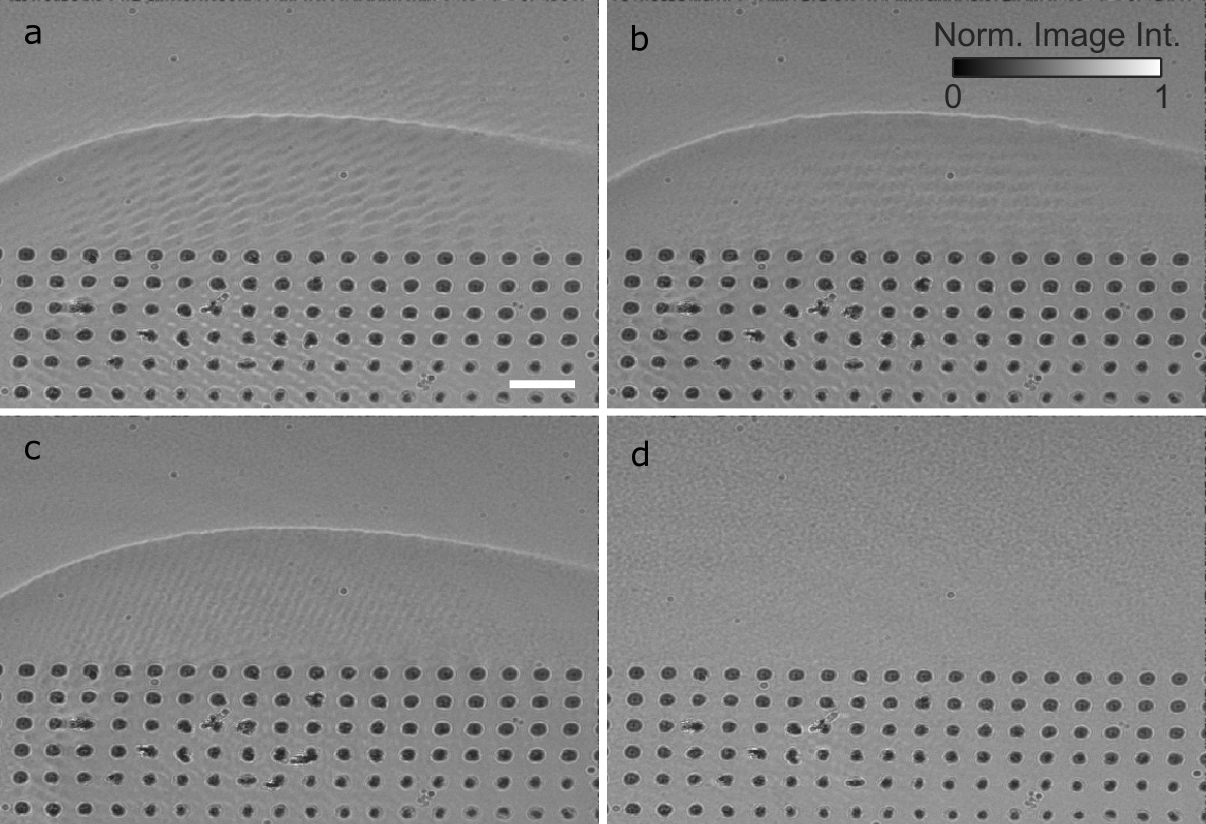}
 \caption{Standing-light-wave-induced charging in a thin film of amorphous carbon deposited by the electron beam. The Lorentz micrographs were recorded in a sample region which was previously exposed to high electron-dose rates for a long duration. Also laser-induced damage of the gold nanoislands in the lower part of the micrographs is observable. The relative angles of the half-waveplate used to set the polarization of the incident light were (a-c): 0°, 40° and 60°, respectively. The illumination is switched off in (d). The light-induced standing-wave patterns were not observable after oxygen-plasma cleaning of the surface or in areas of the sample that were not exposed to a high electron dose. The bright curved line in the upper third of the micrographs marks the area that has been exposed to the high electron dose.}
  \label{fgr:S2}
\end{figure}

\begin{figure}
  \includegraphics[scale=1]{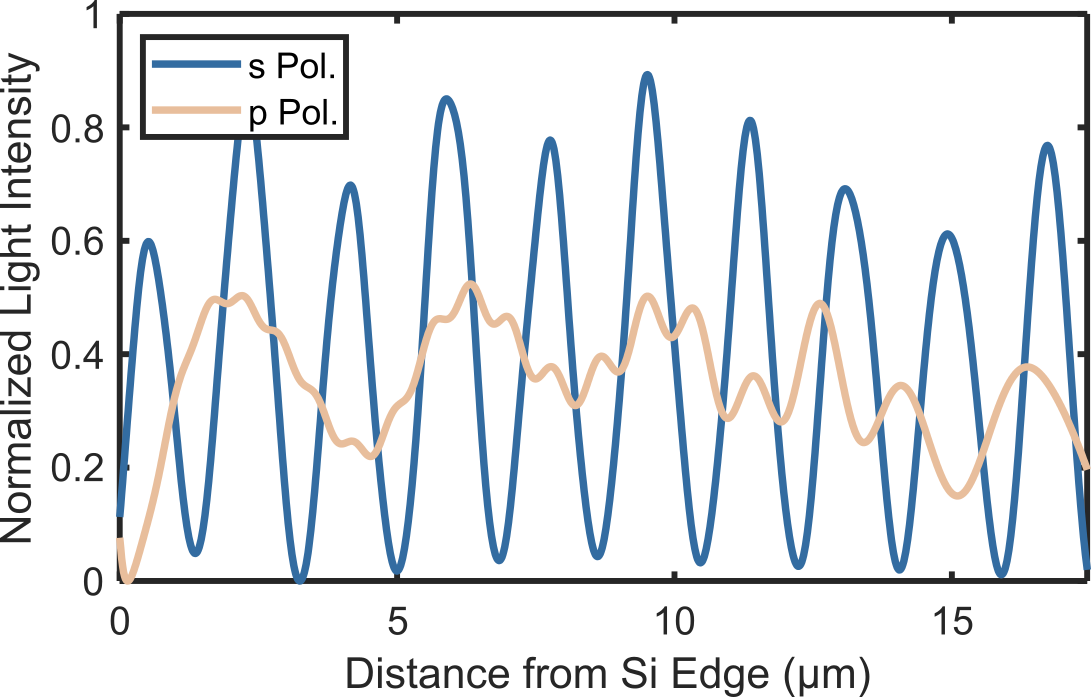}
 \caption{Finite-difference time-domain simulations showing the inhomogeneous distribution of light intensity at the edge of the silicon support frame in a Si$_3$N$_4$~-~Al$_2$O$_3$ heterostructure for s- and p-polarized incident light.}
  \label{fgr:S3}
\end{figure}

\newpage
\noindent \textbf{\Large{References}}
\begin{enumerate}
\item Philipp, H. R. Optical Properties of Silicon Nitride. \textit{J. Electrochem. Soc.} \textbf{1973}, \textit{120} 295.
\item Malitson, I. H. Refraction and Dispersion of Synthetic Sapphire. \textit{J. Opt. Soc. Am.} \textbf{1962}, \textit{52}, 1377
\item Aharonov, Y.; Bohm, D. Significance of Electromagnetic Potentials in the Quantum Theory. \textit{Phys. Rev.} \textbf{1959}, \textit{115}, 485-491.
\item Beleggia, M.; Kasama, T.; Dunin-Borkowski, R. E.; Hofmann, S.; Pozzi, G. Direct measurement of the charge distribution along a biased carbon nanotube bundle using electron holography. \textit{Appl. Phys. Lett.} \textbf{2011}, \textit{98}, 243101.
\item Weber, J. T.; Schäfer, S. Electron Imaging of Nanoscale Charge Distributions Induced by Femtosecond Light Pulses. \textit{Nano Lett.} \textbf{2024}, \textit{24}, 5746-5753
\item McCartney, M.R.; Dunin-Borkowski, R. E.; Smith, D. J. Quantitative measurement of nanoscale electrostatic potentials and charges using off-axis electron holography: Developments and opportunities. \textit{Ultramicroscopy} \textbf{2019}, \textit{203}, 105-118
\end{enumerate}

\end{document}